\documentclass[twocolumn]{aa}
\usepackage{graphicx}
\usepackage{txfonts}
\usepackage[authoryear]{natbib}
\bibpunct{(}{)}{;}{a}{}{,}
\newcommand{\src}  {4U\,0115+63}
\newcommand{\cas} {V635 Cas}
\newcommand{\ha}  {H$\alpha$}
\newcommand{\ew}  {EW(H$\alpha$)}

\def\simless{\mathbin{\lower 3pt\hbox
     {$\rlap{\raise 5pt\hbox{$\char'074$}}\mathchar"7218$}}}   
\def\simmore{\mathbin{\lower 3pt\hbox
     {$\rlap{\raise 5pt\hbox{$\char'076$}}\mathchar"7218$}}}   

\def\msun{~{\rm M}_\odot}
\def\rsun{~{\rm R}_\odot}

\begin{document}

   \title{The Be/X-ray transient 4U\,0115+63/V635 Cassiopeiae \\
III. Quasi-cyclic variability \thanks{Tables 2 and 3 are only available 
in electronic form at the CDS via anonymous ftp to cdsarc.u-strasbg.fr 
(130.79.128.5) or via http://cdsweb.u-strasbg.fr/cgi-bin/qcat?J/A+A/}
}

   \subtitle{}

   \author{
   P. Reig\inst{1,2}
          \and
	  V. Larionov\inst{3}
	  \and
	  I. Negueruela\inst{4}
	  \and
	  A.A. Arkharov\inst{5}
	  \and
	  N.A. Kudryavtseva\inst{3,6}
          }

\authorrunning{Reig et~al.}
\titlerunning{Quasicyclic variability in 4U 0115+63}


   \institute{IESL, Foundation for Reseach and Technology-Hellas, 71110, 
   		Heraklion, Greece
	 \and Physics Department, University of Crete, 71003, 
   		Heraklion, Greece \\
		\email{pau@physics.uoc.gr}
	\and Astronomical Institute, St.Petersburg University, 
198904 St. Petersburg, Russia \\
		\email{vml@vl1104.spb.edu}	 
        \and Departamento de F\'{\i}sica, Ingenier\'{\i}a de Sistemas y Teor\'{\i}a
de la Se\~nal, Universidad de Alicante, E-03080 Alicante, Spain\\
             \email{ignacio@dfists.ua.es}
        \and Pulkovo Central Astronomical Observatory, Pulkovskoe shosse 65, 
	196140 St. Petersburg, Russia \\
	\email{arkadi@arharov.ru}
	\and Max-Planck-Institut fur Radioastronomie, Auf dem Huegel 69, 
	Bonn 53121, Germany
 }

   \date{Received ; accepted}

\abstract
{
\src\ is one of the most active and best studied Be/X-ray transients. 
Previous studies of \src\ have led to the suggestion that \src\ undergoes 
relatively fast quasi-cyclic activity. However, due to the lack of good
coverage of the observations, the variability time scales are uncertain.
}
{Our objective is to investigate the long-term behaviour of \src/\cas\ to 
confirm its quasi-cyclic nature and to explain its correlated optical/IR and
X-ray variability.}
{We have performed optical/IR photometric observations and optical 
spectroscopic observations of \src/\cas\ over the last decade with unprecedented 
coverage. We have focused on the \ha\ line variability and the 
long-term changes of the photometric magnitudes and colours and
investigated
these changes in correlation with the X-ray activity of the source.}
{The optical and infrared emission is characterised by cyclic changes with 
a period of $\sim$ 5 years.  This long-term variability is attributed to 
the state of the circumstellar disc around the Be star companion. 
Each cycle involves a low state when the disc is very weak or absent
and the associated low amplitude variability is orbitally modulated and 
a high state when a perturbed disc precesses, giving rise to fast and large 
amplitude photometric changes. X-ray outbursts in \src\ come in pairs, i.e., 
two in every cycle. However, sometimes the second outburst is missing.
}
{Our results can be explained within the framework of the decretion disc model.
The neutron star acts as the perturbing body, truncating and distorting 
the disc.  The first outburst would occur before the disc is strongly
perturbed. The second outburst leads to the dispersal of the disc and marks 
the end of the perturbed phase.
}


   \maketitle
\begin{table*}
\begin{center}
\caption{Historical record of the X-ray outbursts and spin period of \src.}
\label{his}
\begin{tabular}{clcl||cll}
\hline \hline \noalign{\smallskip}
Date	&Satellite	&Intensity	&Reference	&MJD	&$P_{\rm spin}$	&Reference\\
	&		&(Crab)		&	\\
\hline \noalign{\smallskip}
1969 Aug  &Vela 5B		&0.38		&\citet{whit89}	&40963  & 3.614658$\pm$0.000036	&\citet{kell81}   \\
1970 Jan  &Vela 5B		&0.23		&\citet{whit89}	&42283  & 3.6142$\pm$0.0001  	&\citet{whit89}	 \\
1970 Aug  &Vela 5B		&0.22		&\citet{whit89}	&43556  & 3.61355$\pm$0.000000 	&\citet{john78} \\
1971 Jan  &Uhuru		&0.17		&\citet{form76}	&44589  & 3.6146643$\pm$0.0000018 &\citet{rick81} \\
1974 Aug  &Vela 5B		&1.8		&\citet{whit89}	&47932  & 3.614690$\pm$0.000002 &\citet{tamu92} \\
1978 Jan  &SAS 3		&1.7		&\citet{comi78}	&47942  & 3.61461$\pm$0.00001  	&\citet{luto00}	 \\
	  &Ariel 5		&0.5		&\citet{rose79} &49481  & 3.6145107$\pm$0.0000010&\citet{scot94} \\
	  &HEAO 1		&0.96		&\citet{john78} &50042  & 3.614499$\pm$0.000004 &\citet{fing95}  \\
1980 Dec  &Ariel 6		&0.18		&\citet{rick81}	&50307  & 3.614452$\pm$0.000003 &\citet{scot96}  \\
1987 Feb  &Ginga		&0.18		&\citet{tsun88}	&51232  & 3.61452$\pm$0.00001 	&\citet{wils99}	 \\
1990 Feb  &Ginga		&0.4		&\citet{tamu92}	&53261  & 3.614$\pm$0.003  	&\citet{zuri04}	 \\
	  &GRANAT		&0.8		&\citet{luto00} &53266  & 3.616$\pm$0.001 	&\citet{tuel04}	 \\
1991 Apr  &CGRO			&0.08		&\citet{comi94}	&&&	\\
1994 May  &CGRO			&0.18		&\citet{wils94}	&&&	\\
1995 Nov  &CGRO			&0.07		&\citet{fing95}	&&&	\\
	  &GRANAT		&0.7		&\citet{sazo95} &&&	\\
1999 Mar  &RXTE			&0.4		&This work	&&&	\\
	  &BeppoSAX		&0.3		&\citet{sant99}	&&&	\\
	  &CGRO			&$>$0.07	&\citet{wils99}	&&&	\\
2000 Sep  &RXTE			&0.2		&This work	&&&	\\
2004 Sep  &RXTE			&0.3		&This work	&&&	\\
	  &INTEGRAL		&0.4		&\citet{zuri04}	&&&	\\
\hline \hline \noalign{\smallskip}
\multicolumn{7}{l}{Vela 5B: 3-12 keV, Uhuru: 2-6 keV, HEAO 1: 3-13 keV, 
SAS 3: 1-27 keV, Ariel 5: 3-6 keV, Ariel 6: 1-50 keV,Ginga: 1-20 keV}\\
\multicolumn{7}{l}{CGRO: 20-50 keV,GRANAT: 8-20 keV, 
RXTE: 2-12 keV, BeppoSAX: 2-10 keV, INTEGRAL: 20-60 keV}\\
\end{tabular}
\end{center}
\end{table*}

\section{Introduction}

\src\ was one of the first Be/X-ray binaries to be discovered. The oldest
available X-ray observation dates back to August 1969 when the {\em Vela
5B} satellite detected the source as three small outbursts separated by 180
days \citep{whit89}. Since then about 15 outbursts have been reported (see
Table~\ref{his}). Normally, these outbursts represent an increase in the
X-ray luminosity by a factor $\sim$ 100 and last for about a month. The
strongest outbursts reach luminosities close to the Eddington value ($\sim
10^{38}$ erg cm$^{-2}$ s$^{-1}$).  The lack of orbital modulation and the
relative large increase in luminosity define these outbursts as type II. 
Only in one occasion, in 1996, has \src\ showed orbitally modulated,
shorter and smaller outbursts, i.e., type I \citep{negu98}. For a review of
the X-ray variability in Be/X-ray binaries see, e.g., \citet{coe00} and
\citet{ziol02}.

X-ray pulsations ($P_{\rm spin}=3.6$ s) were soon discovered \citep{comi78}
and the orbital parameters ($P_{\rm orb}=24.3$ d, $e=0.34$, $a_x \sin
i=140.1$ lt-s) determined \citep{rapp78}.  Subsequent detections led to the
discovery of one \citep{whea79}, two \citep{white83}, three \citep{hein99}
and four \citep{sant99} cyclotron resonance scattering features.

The 1978 outburst allowed the identification of the optical counterpart
\citep{john78, hutc81} with a reddened ($A_V \simmore 5$ mag)  B-type star
that showed \ha\ in emission, \cas. Extensive studies in the IR/optical
\citep{kris83,mend91,unge98,negu01}  and X-ray \citep{whit89,tsun88} bands
allowed the determination of the astrophysical parameters --- \cas\  is a
$V \approx 15$ B0.2Ve star located at a distance of $\sim$ 7-8 kpc --- and
to the suggestion of cyclic changes with a quasi-period of 3--5 years. This
quasi-cycling behaviour is closely related to the dynamical evolution of
the viscous circumstellar disc around the Be star. The Be star loses and
reforms the disc on time scales of 3-5 years \citep{negu01}. At some point
during the growing phase the disc becomes unstable and highly disturbed.

In this paper we present the most complete monitoring of \src/\cas\ in the
optical and infrared wavelength bands made up to now. Our observations are
used in combination with published data to investigate its quasi-cyclic
optical/IR/X-ray variability. The results presented here build on previous
work by \citep[][Paper~I]{neok01} and \citet[][Paper~II]{negu01}. 

\section{Observations}

\subsection{Optical and infrared photometry}

Optical ($BVRI$) photometric observations were obtained from the 20cm
(before April 2001) and 70cm (after April 2001) telescopes at the Crimean
Observatory (CRI) in Ukraine between 1998-2005 using an ST-7 CCD of the
St.Petersburg University. The array is 765 $\times$ 510 pixels,
corresponding to a $8.1\arcmin\times5.4\arcmin$ field of view
($0.65\arcsec$/pixel scale). The data set consists of about 260 $B$ and 350
$VRI$ measurements. In order to improve the signal-to-noise ratio, up to 5 
images in each colour band were obtained and co-added to create the final
image. The standard technique of bias and dark subtraction and
flat-fielding was used. A calibrated set of standard stars, located within
the same field, was used to perform aperture photometry using a
SExtractor-based package. The photometric accuracy is usually better than
$0\fm01$ in $V$, $R$ and $I$ and about $0\fm03$ in $B$. 

Near-infrared photometric data (about 200 $JHK $measurements) were obtained
at Campo Imperatore (Italy) with SWIRCAM  NIR camera with the PICNIC
$256\times256$ pixel array, attached to the 1.1~m telescope. The camera's
field of view is $4.5\arcmin\times4.5\arcmin$ with a $1.04\arcsec$/pixel
scale.  Each photometric image was obtained from 5 co-added dithered
images, after sky subtraction and flat-field correction. The same
photometric package as for the ST-7 data was used to perform photometry. 

Table 2 (electronic form only) gives the results of our photometric
monitoring.

\subsection{Optical spectroscopy}

The main source of spectroscopy is the 1.3~m Skinakas telescope (SKI),
which was equipped with a $2000\times800$ ISA SITe
CCD and a 1302 l~mm$^{-1}$ grating, giving a nominal dispersion of 1.04
\AA/pixel. Many other spectra were obtained through the service programme
of the Isaac Newton Group at La Palma, either with the 2.5~m Isaac Newton
Telescope (INT) or the 4.2~m William Herschel Telescope (WHT). The INT was
equipped with the Intermediate Dispersion Spectrograph (IDS), fit with the
235-mm camera and different intermediate-resolution gratings. The WHT
spectra were obtained with the red arm of the Intermediate Dispersion 
Spectroscopic and Imaging System (ISIS), equipped with either the R600R or
R1200R gratings.

Other observations have been obtained with the 2.6~m Nordic Optical
Telescope (NOT), also located in La Palma, equipped with ALFOSC; the 1.93~m
of the Haute  Provence observatory (OHP) in France, equipped with the {\it
Car\'elec} spectrograph and the Mt. Ekar 1.82~m Telescope (EKA) of the
Padova Astronomical Observatory (Italy), equipped with AFOSC. One low
resolution spectrum was taken with the blue arm of the TWIN spectrograph
on  the 3.5~m of the Calar Alto observatory (CA) in Spain. Table~3
(electronic form only) gives the log of the optical spectroscopic
observations. 

The reduction of the spectra was made using the STARLINK {\em
Figaro} package \citep{shor01}, while their analysis was performed using the
STARLINK {\em Dipso} package \citep{howa98}. 

\begin{figure}
\begin{tabular}{c}
\resizebox{\hsize}{!}{\includegraphics{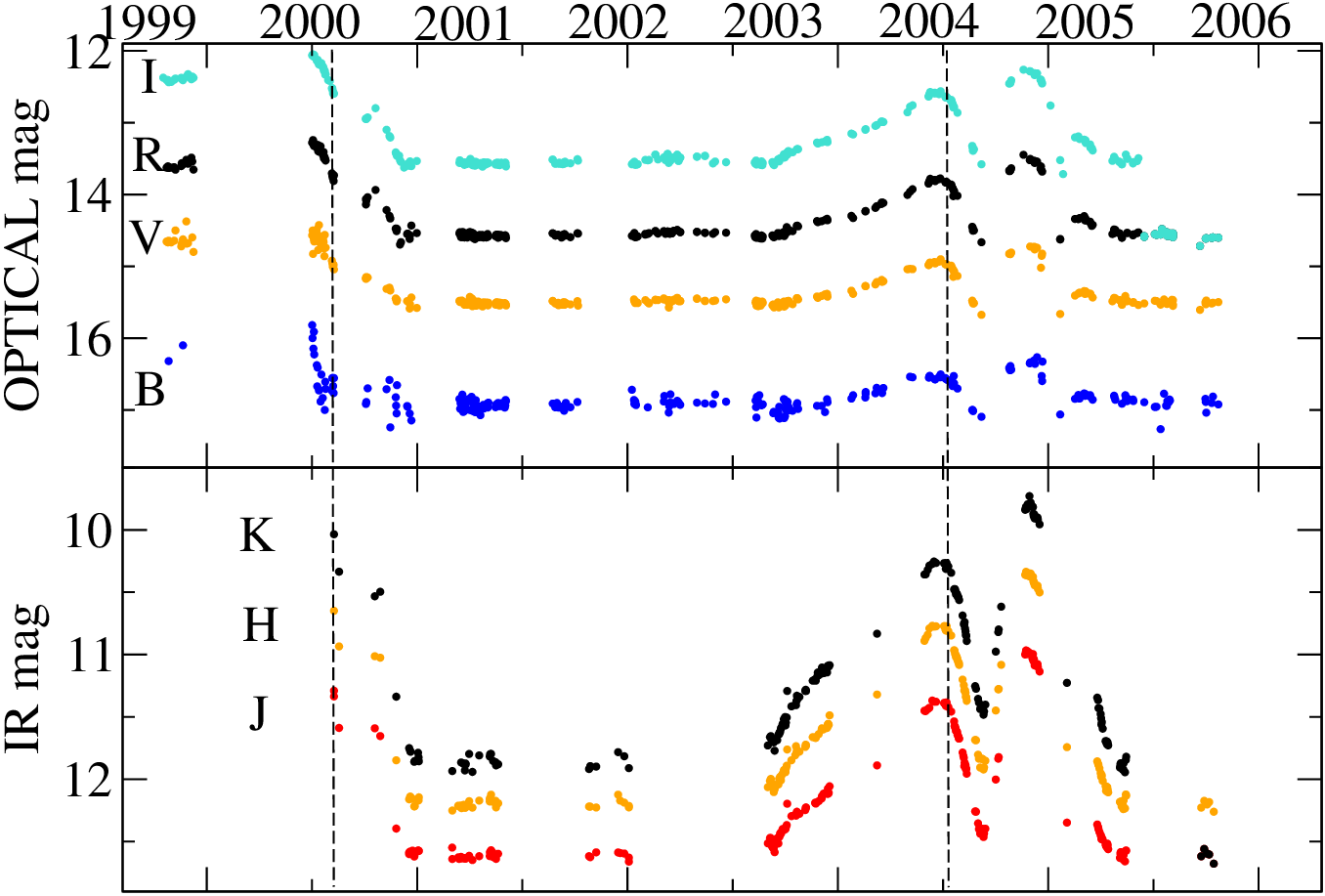}} \\
\resizebox{\hsize}{!}{\includegraphics{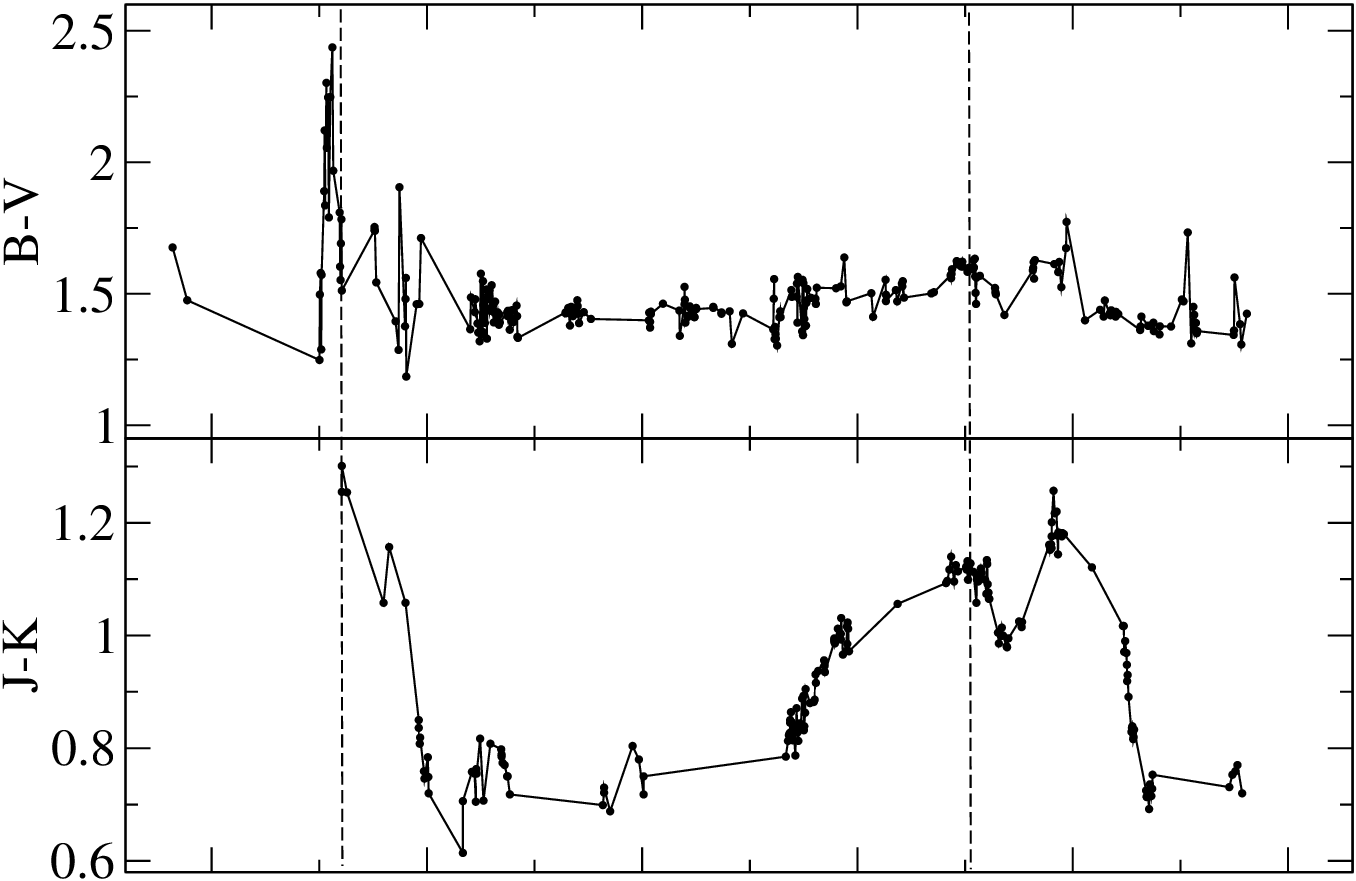}} \\
\resizebox{\hsize}{!}{\includegraphics{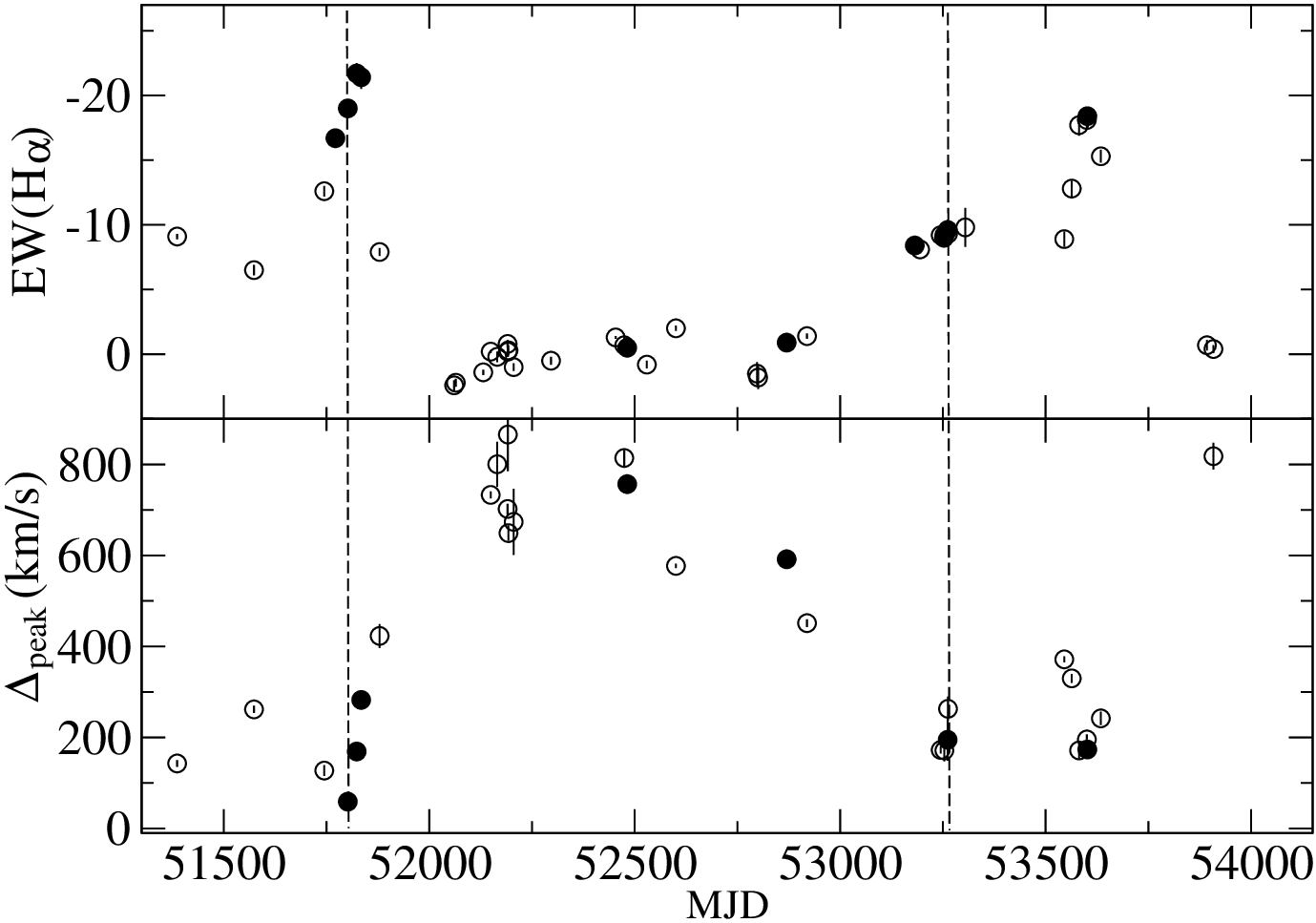}} \\
\end{tabular}
\caption[]{Long-term evolution of the optical and infrared magnitudes and
colours and of the \ha\ equivalent width and peak separation of double-peak
profiles. Dotted lines mark the occurrence of X-ray outbursts. Filled circles 
in the bottom panels correspond to the observations shown in Fig.~\ref{cycle}}
\label{phot}
\end{figure}

\section{Results}

We have been monitoring \src\  spectroscopically since the early 1990's
(Paper~II). Photometric data were also occasionally acquired. Since August
1999 the source is being monitored in the $UBVRIJHK$ bands. In this section
we present the results of these new observations. The analysis of the
historical variability curves will be presented in Sect.~\ref{discussion}.

\subsection{Photometric variability}
\label{pho}

Fig.~\ref{phot} shows the evolution of the optical and infrared magnitudes
and colours for the period August 2000--June 2006. These observations
define a complete cycle of variability. The light curve begins with the
photometric bright state that led to the 2000 X-ray outburst.  The
magnitudes gradually decreased and the colours became bluer. A faint stable
photometric state was reached at around MJD 52000 (April 2001). This state
extends for more than two years up to $\sim$ MJD 52880 (August 2003). We
will refer to this state as the extended photometric low state (ELS).   The
average $BVRI$ magnitudes during the ELS are $B=16.93\pm0.05$,
$V=15.50\pm0.03$, $R=14.56\pm0.03$, $I=13.55\pm0.04$, $J=12.61\pm0.03$,
$H=12.19\pm0.03$, $K=11.86\pm0.05$. These values are in complete agreement
with those given in Paper~I for the previous faint state. Photometrically,
the ELS is a quiet state with very low  amplitude variations, as can be
deduced from the small values of the standard deviation. 

The end of the ELS is marked by the gradual brightening of the photometric
bands in August 2003. The source enters a new activity state
characterised by large amplitude variations in the form of optical
eruptions, i.e., the optical/IR brightness alternates between  maxima and
ELS values. The amplitude of variability is larger at longer wavelengths
(e.g. $V$ increased by $\sim$0.5 mag and $K$ by 1.6 mag in about 440 days).
In coincidence with the first optical eruption of this state, a new X-ray
outburst was observed (September 2004). Note also the constancy of the
$(B-V)$ colour throughout the observations. 

\citet{bayk05} reported ROTSE observations covering the 2004 outburst. They
found a sharp drop lasting for about a week (MJD 53235--53242) of $\sim0.3$
mag in the source brightness a few days before the onset of the X-ray
outburst and interpreted this result as a sign of mass ejection from the
outer parts of the disc of the Be star. We do not observe such drop in our
data. Although our observations contain only one point in the 
narrow interval when the ROTSE drop was seen, it shows very
similar values to those of the previous and following observations.

\subsection{Spectroscopic variability}

Figure~\ref{phot} also shows the evolution of the equivalent width (\ew)
and  the separation of the peaks, in km s$^{-1}$, of the split profiles 
of the \ha\ line in the interval 1999-2006. As the \ew\ increases the peak
separation decreases. This is the expected behaviour of a quasi-Keplerian
disk and indicates that as the \ew\ increases, the region where the \ha\
line is produced moves further away from the central star
\citep[e.g.][]{humm95}.

\begin{figure*}
\resizebox{\hsize}{!}{\includegraphics{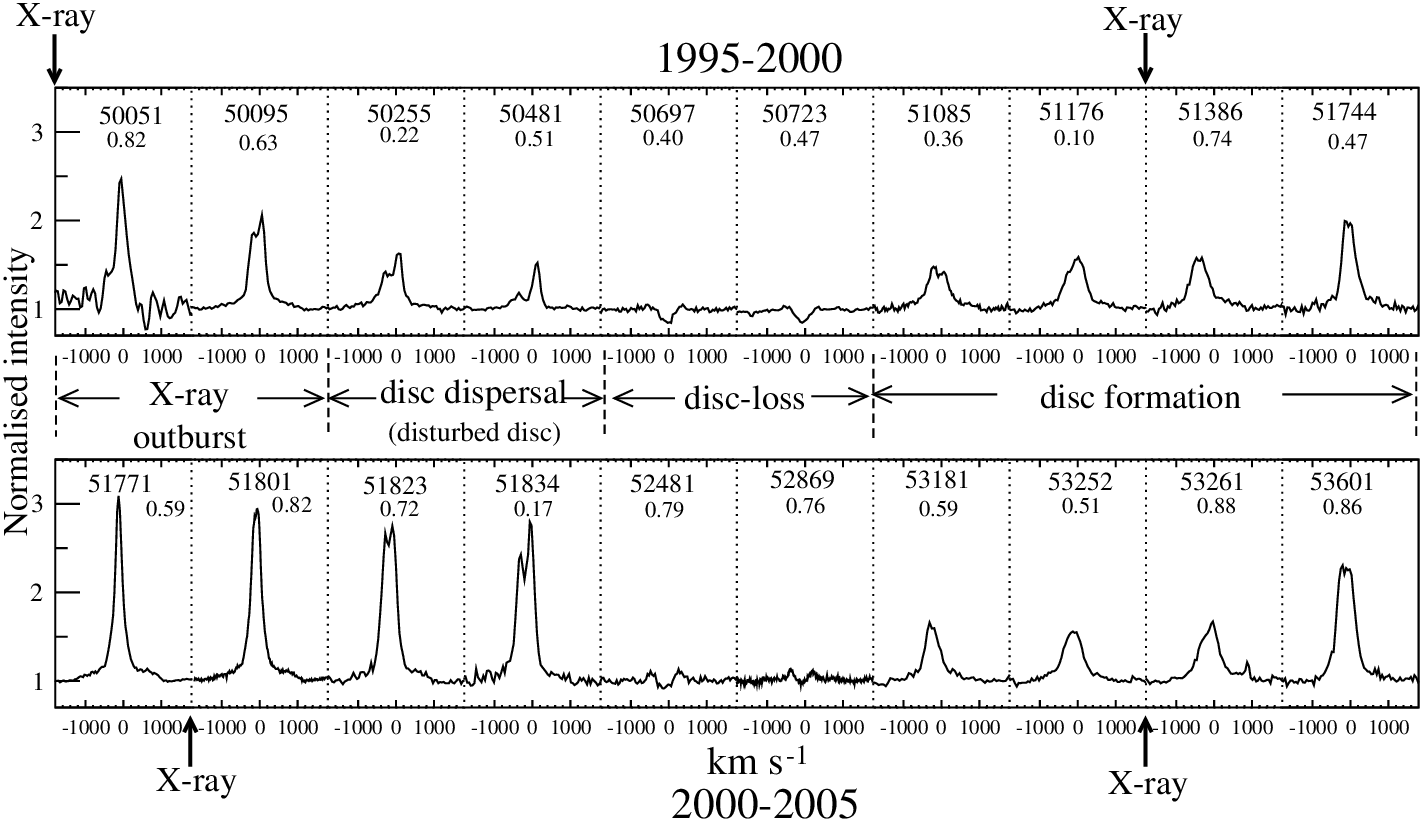}}
\caption[]{Evolution of the H$\alpha$ line profile over the $\sim$ 5 year 
quasi-cycle.  The spectra were normalised to the neighbouring continuum
and the wavelength converted to velocity units. 
Indicated are the MJD and orbital phase according to the orbital solution
of \citet{tamu92}. The Y-axis scale was left the
same in all panels to facilitate comparison.}
\label{cycle}
\end{figure*}

In Fig.~\ref{cycle} the \ha\
profile  through the different phases of the variability cycle is
displayed. The top panel corresponds to the cycle that began after the 1995
outburst (Paper~II), while the bottom panel to that after the 2000
outburst.  The emission line profile of the \ha\ line shows significant
richness in variability. Both single- and double-peak as well as symmetric
and asymmetric profiles are seen. Symmetric profiles are associated with
extreme values, either way, of the \ew. Asymmetric profiles are associated
with intermediate values of \ew.

Taking as the starting point of the cycle the X-ray outburst prior to the
ELS, the evolution of the \ha\ profile through the cycle is the following:
at the time of the outburst the \ha\ line shows a single peak profile. The
\ew\ finds itself in a relative maximum. When the profile is a narrow
single peak there is a broad base component on top of which it sits (see
e.g. Fig.~\ref{cycle} spectra MJD 51771 and MJD 51801). This component was
already noticed in \citet{negu01} and may be interpreted as evidence for a
warped disc. The broad base presumably represents the motion in the inner
part of the disk, where the velocity components are higher. After the
X-ray outburst the \ha\ profile becomes double peaked and asymmetric with
the strength of the red peak larger than that of the blue peak (V$<$R
phase).  The \ew\ decreases gradually until the source enters the ELS. At
this point, \ew\ presents its minimum value. Absorption as well as
double-peak profiles are seen. For some spectra the depression between the
double-peak profile extends below the stellar continuum. As the optical
activity increases, the profile shows a distorted structure again, although
not as marked as during the decay. At the beginning of this phase
single-peak and red-dominated double-peak profiles are present. As the \ew\
increases blue-dominated double-peak profiles start to appear. When the
\ew\ reaches $\sim$10 \AA\ a new X-ray outburst takes place. This outburst
does not alter the strength of the \ha\ line significantly, although the
overall brightness of the source decreases. By August 2005, the strength of
the \ha\ line and the IR emission had experienced a fast and sudden
increase. The \ew\ almost doubled, while the IR magnitudes became $\sim$ 1
mag brighter, on time scales of a few tens of days. It is as though the
source was preparing itself for a new X-ray outburst. However, our latest
observations (June 2006) show, both the \ew\ and photometric magnitudes,
close to ELS values and no evidence of renewed X-ray activity.

\subsection{Reddening and distance}

\begin{figure}
\resizebox{\hsize}{!}{\includegraphics{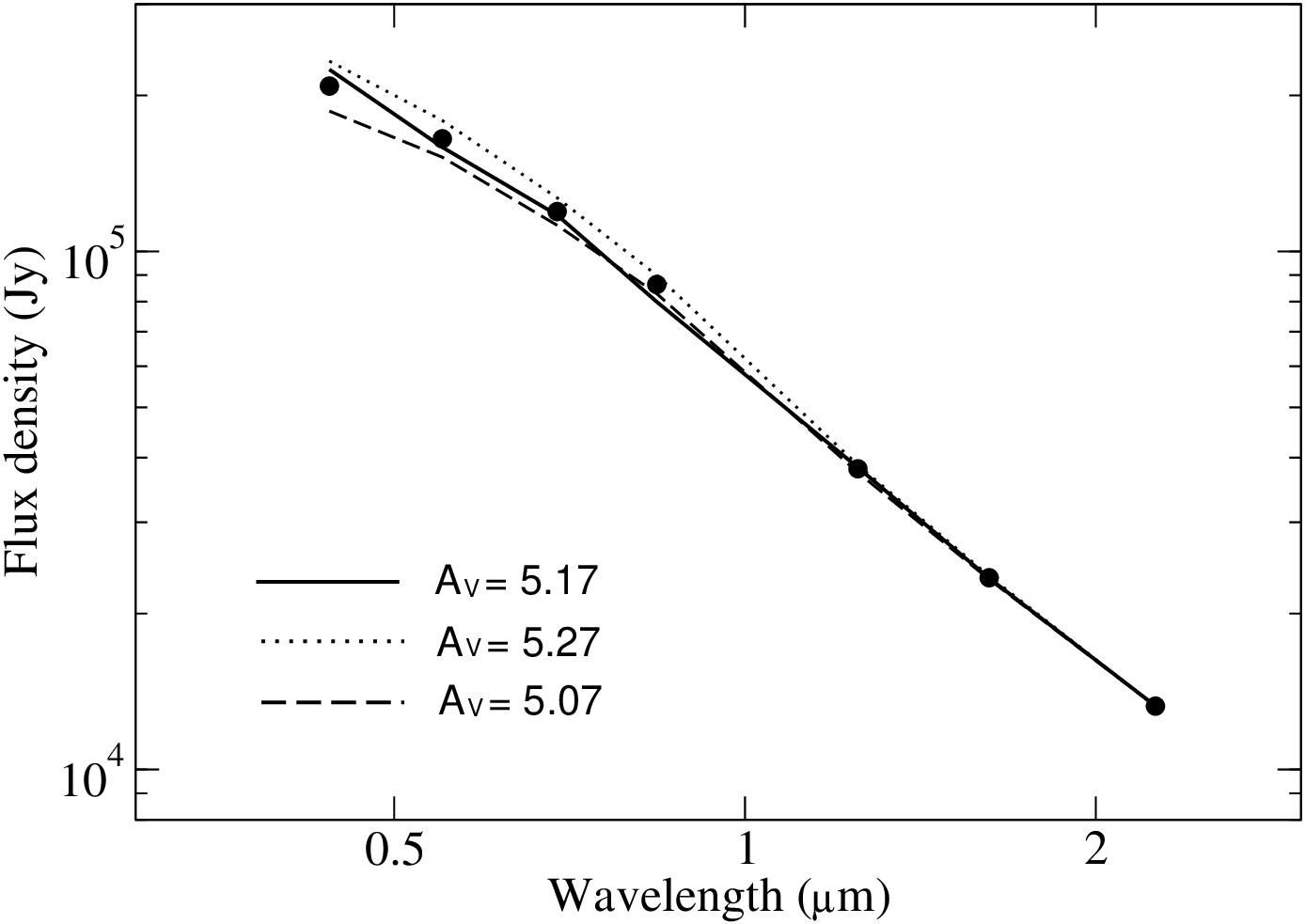}}
\caption[]{Energy distribution of a B0V star dereddened using $A_V=5.17$. }
\label{sed}
\end{figure}

In Paper~I values of the reddening and distance were derived from the
observations showing the bluest colours (those from January 8, 1998). It
was then   assumed that these observations corresponded to purely
photospheric emission. In view of the observations presented here, a
revision of these parameters is justified since {\em i)} a much longer set
of optical photometric observations is available,  {\em ii)} a longer
extended low optical state (MJD 52000-52800) is observed, where presumably
the underlying B star is exposed and photospheric emission without
substantial  contribution from the disc is detected and {\em iii)}  the
observations cover a wider wavelength band with (quasi)simultaneous optical
and near IR data, which allows us to estimate the reddening by
fitting the photometric data to a model atmosphere. Fig.~\ref{sed} shows
the energy distribution of a B0V star \citep{stra95} and the lowest
magnitudes of the ELS. Letting $A_V$ be a free parameter we find that the
best fit is achieved for $A_V=5.17\pm0.03$. The distributions with
$A_V=\pm3\sigma$ are also shown.

Taking into account that $A_K = 0.115A_V$, that for a B0V the intrinsic
$K_0=-3.25$ \citep{stra95} and the minimum observed magnitude,
$K=11.89\pm0.02$, the distance-modulus is $DM=K-0.115A_V-K_0=14.55$ or 
$d=8.1\pm 0.1$ kpc. The error in the distance includes the photometric
uncertainty only and assumes that neither $K_0$ nor the reddening law are
affected by errors --- for example, an uncertainty of 0.2 mag in $K_0$
translates into a distance error of about 0.7 kpc.

The assumption of standard reddening may introduce large errors if not
verified. In Paper I, it was shown that the use of the extinction law by
\citet{fitz99} favours a value of $R$ close to the standard $R=3.1$. For
further verification, we used the {\sc chorizos} code \citep{maiz04} to study
the reddening. The program was used to fit extinction laws from
\citet{card89} convoluted with the spectral energy distribution of a $T_{{\rm
eff}}=27,500\:{\rm K}$,$\log g =4.0$ {\sc tlusty} atmosphere model to our
photometry. Although we obtain a slightly lower value for $R$, the derived
$A_{V}=4.8\pm0.1$ is compatible with the value found above at a $3\sigma$
level.


\subsection{Power spectrum analysis}
\label{pow}

In order to search for periodic variations in the photometric light curves
we performed a power spectral analysis. The details of the technique
employed can be found in \citet{lari01}.

The power spectrum and spectral window of the data set were calculated as

\begin{equation}
P(f)=\frac{1}{N^2}\left | \sum\limits_{j=1}^N
(m_j-\overline{m})\mathrm e^{-\mathrm{i} 2\pi f t_j}\right |^2
\label{eq:ps} \end{equation}

and

\begin{equation}
W(f)=\frac{1}{N^2}\left | \sum\limits_{j=1}^N
\mathrm e^{-\mathrm{i} 2\pi f t_j}\right |^2, \quad
W(0)=1, \label{eq:w} \end{equation}

\noindent respectively, where $\mathrm{i}=\sqrt{-1}$,  $t_j$
denotes the time of observation, and $m_j$ and $\overline{m}$,
the individual and mean values of the brightness, respectively.

It is clear from Fig.~\ref{phot} that in order to judge with
sufficient confidence the reality of the orbital plus close and/or
related periodicities, one should remove the longer-period
component(s) of variability. In our case it is not sufficient to
subtract the linear trend, and we need to take into account the
slow light variations. We have constructed a smoothed data
set using the method of a sliding mean with the window value
$\Delta $, replacing raw data $m_\mathrm{i}$ for each time
$t_\mathrm{i}$ by the weighted mean:

\begin{equation}
m_i^{\prime}= -2.5\log\left(\frac 1{\sum p_j}
\sum\limits_{j=1}^k p_j \cdot 10^{-0.4m_j}\right), \label{eq:1}
\end{equation}

\noindent where $k$ is the number of data points within interval $
[t_\mathrm{i}-\Delta, t_\mathrm{i}+\Delta]$, and the weight of
$j$th point is determined as

\begin{equation}
p_\mathrm{j}=\exp\left[-(\delta t_\mathrm{j}/\Delta)^2\right],
\label{eq:2}
\end{equation}

\noindent where $\delta t_\mathrm{j}$ is the time span from the
$j$th point to the center of the window. The optimal value of the
smoothing interval was searched by trial and error within a range
$10^\mathrm{d}< \Delta < 30^\mathrm{d}$. The criterion for
$\Delta$ selection was the signal-to-noise ratio of the
power-spectrum peaks in the region of interest, i.e. around the
orbital frequency.

We divided the photometric light curves into several intervals and analysed
them separately: MJD~51400--52000 corresponds to the brighter optical state
of the source prior to the ELS, MJD ~52100--52850 corresponds to the
ELS, MJD~52850--53250 includes the smooth brightening after the ELS up to
the 2004 X-ray outburst and MJD~53250--53715, which covers the time
interval after the X-ray outburst.

\begin{figure}
\resizebox{\hsize}{!}{\includegraphics{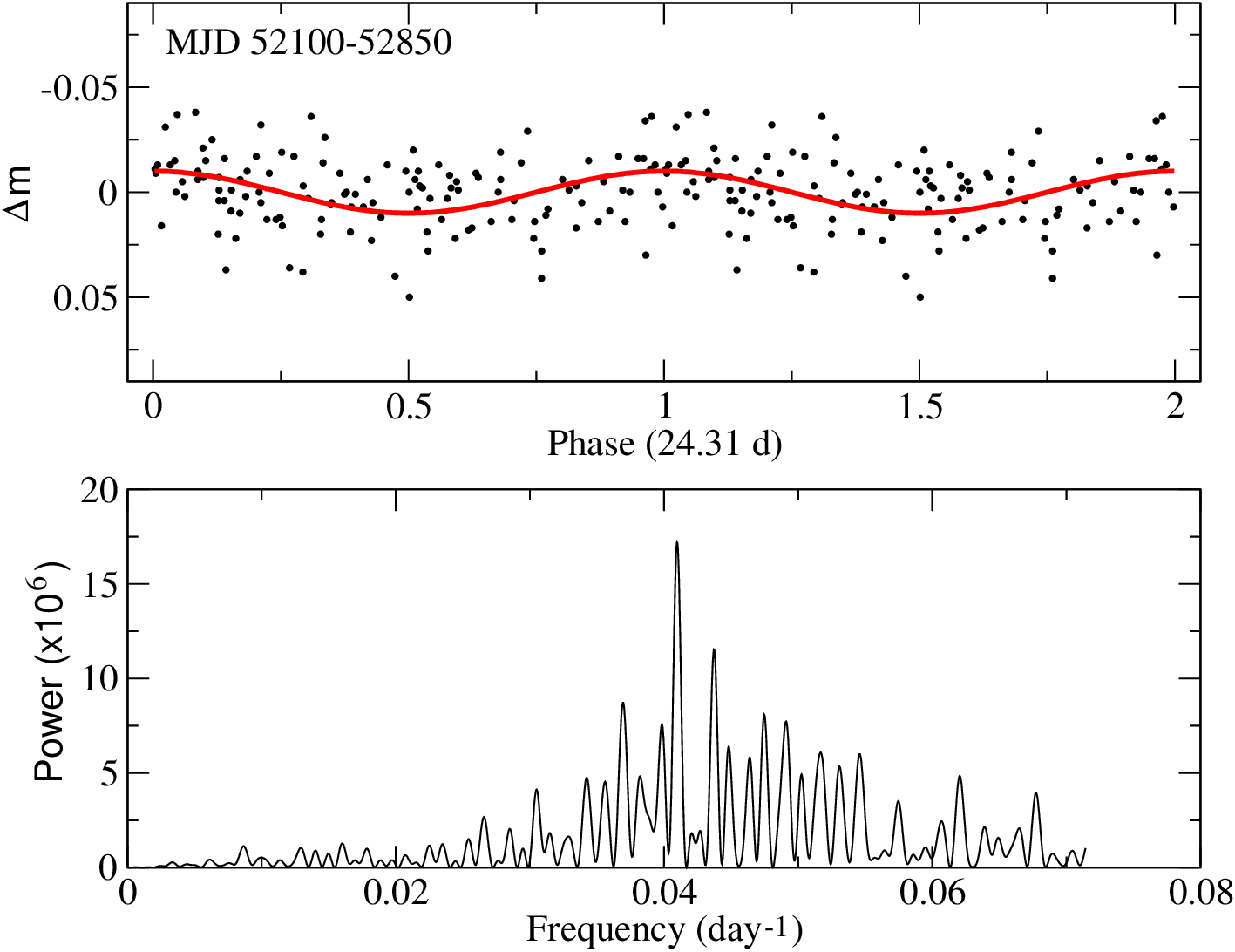}}
\caption[]{{\it Top:} R-R$^\prime$ dependence on the phase of the 24.31-day
period for the ELS.
{\it Bottom:} Power spectrum of the residuals $R-R^\prime$ obtained from
the period MJD 52100--52300. The most prominent peak corresponds to 
$P=24.41$ days. }
\label{powsp}
\end{figure}

A clear modulation was found during the ELS, independent of the value of
the smoothing interval adopted. The most prominent peak in the power
spectrum corresponds to a period of $24.4\pm0.1$ d, which is consistent
with the orbital period. The
low-frequency components are most effectively suppressed and the highest
signal-to-noise ratio is obtained when $\Delta = 15^\mathrm{d}$. The power
spectrum for the residuals $R_\mathrm{j}-R_\mathrm{j}^{\prime}$ is shown in
Fig.~\ref{powsp}.

The probability of chance occurrence of a peak with amplitude
$P_\mathrm{max}$ in the power spectrum with a mean value $P_\mathrm{mean}$ 
can be estimated as $\Phi =100\%\cdot \{1-[1-\exp
[-(P_\mathrm{max}/P_\mathrm{mean})]]^{N_\mathrm{ind}}\}$; for non-uniformly
spaced data  the number of independent frequencies
$N_\mathrm{ind}=-6.362+1.193 \cdot N+0.00098 \cdot N^{2}$, where $N$ is the
number of observations \citep{horn86}. In our case $N=301$ and
$N_\mathrm{ind}=389$; after subtraction of the slow component as described
above (Eqs.~\ref{eq:1} and \ref{eq:2}), $\Phi <0.2\%$.

The amplitude of the sine-wave obtained is $0.01$ mag in $R$. We calculate
the ephemeris of the small-scale optical variations as
$\mbox{JD}_\mathrm{Max} = 2449498(\pm 1.0) + 24.41(\pm0.15)\cdot E$,  where
$E$ is the epoch number. The most striking (but not unexpected) result is
that these values practically coincide, within errors quoted, with the
X-ray ephemeris. The position of the optical maximum corresponds to the
periastron passage.  The orbital period modulation is also present in the
time interval MJD~52850--53250,  but the shape is markedly non-sinusoidal.

No evidence of a 24-day peak was found in the interval MJD 51400-52000.
Instead, there is a modulation with 22.15 day period and amplitude
$\approx0.03$.  One possible explanation for the 22.15-d period is that it
represents the beat frequency between the orbital frequency and precession
frequency of the disc. In this case the disc would precess with retrograde
motion with a period of about 250 days. 
After the 2004 X-ray outburst all periodicities are suppressed.

We also applied the same analysis to the $V$ and $I$ bands. For the $I$ band
we obtained practically the same results as in $R$, while in $V$ we
were not able to detect any significant modulation close to the orbital
period. The fact that the beat frequency is more apparent
in the redder bands gives support to the interpretation of the 22.15-d
periodicity as being affected by the precession of the disc, as one would
expect the IR magnitudes to be more affected  by the disc that the shorter
wavelengths.


\section{Discussion}
\label{discussion}

\subsection{Source states}

We have analysed optical light curves and spectra of \src\ spanning over more
than two decades, the last 10 years with unprecedented coverage. During
this period the source exhibited significant photometric and spectroscopic
variability, allowing the definition of source states. The intensity of the
optical, infrared and X-ray emission highly correlates with the state of
the source.

\begin{figure}
\begin{tabular}{c}
\resizebox{\hsize}{!}{\includegraphics{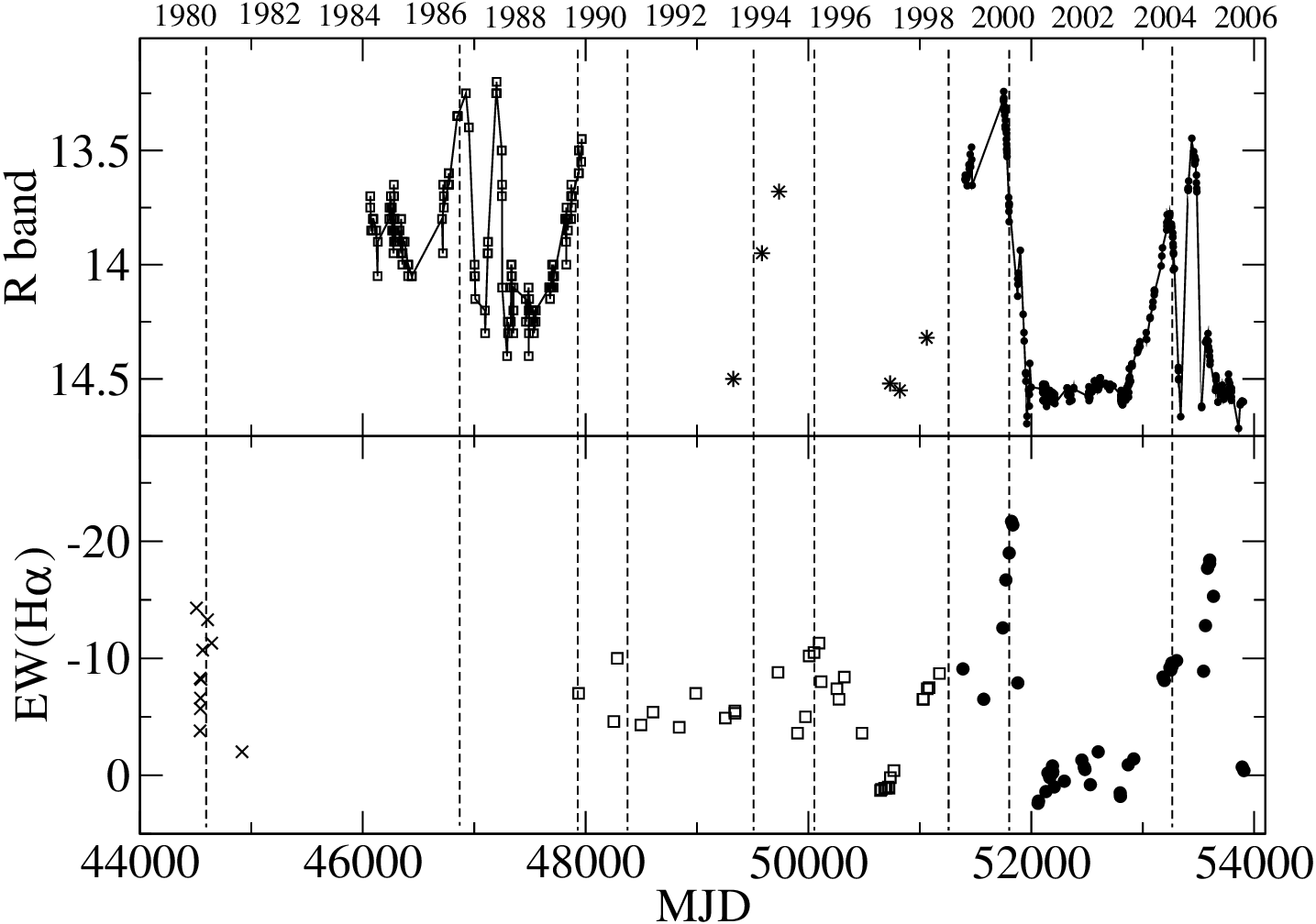}} \\
\end{tabular}
\caption[]{Long-term evolution of the $R$ band and the \ew. 
In the top panel squares are from \citet{mend91} and stars from
\citet{negu01}. In the bottom panel squares are from 
\citet{negu01} and crosses from \citet{kris83}. The data from the year 2000 
onward represent new observations. Dotted lines mark the occurrence 
of X-ray outbursts.}
\label{long}
\end{figure}

Figure \ref{long} shows the historical record (1980-2006) of the $R$ band
and \ha\ equivalent width (\ew) of \cas.  Squares and stars in the top
panel of Fig.~\ref{long} represent observations taken from \cite{mend91}
and \citet{negu01}, respectively. Squares in the \ew\ panel are from
\citet{negu01} and crosses from \citet{kris83}. Circles correspond to our
new measurements.   Vertical dashed lines denote the occurrence of X-ray
outbursts.  Note the striking repeatability of the pattern of variability
during the periods 1986-1990 and 2002-2005: a photometric faint sate is
followed by an optical flare with a slow gradual rise (200--300 days)
and a fast decay (70--80 days). An X-ray outburst occurs when the source is
in a photometric bright state. A second optical eruption separated from the
previous one by about 250 days is not accompanied by X-ray activity. This
second optical eruption presents a more symmetric profile with a fast rise
and equally fast decay. \citet{mend91} also reported the presence of a
small flare after the second eruption, peaking in June 1988 (MJD 47340)
with an amplitude of $\sim$ 0.3 mag and duration of 50 days. Exactly the
same event is seen 17 years after (MJD 53585). If the source follows the
same pattern as in the late 1980's, then we should expect the source to
remain in a low photometric state until spring 2007 and to start a new
optical outburst soon after. The overall difference of $\sim$ 0.25 mag
between the Mendelson \& Mazeh's and our data set is most likely due to
instrumental as well as calibration (selection of different secondary
standard stars) effects, as no attempt to perform absolute photometry was
made in \citet{mend91}. The optical/IR/X-ray behaviour of \src/\cas\  can
be understood in terms of the evolution of the equatorial disc around the 
Be star. Two basic states can be distinguished depending on the presence or
absence or the disc. 

\subsubsection{The low state}

\begin{figure}
\resizebox{\hsize}{!}{\includegraphics{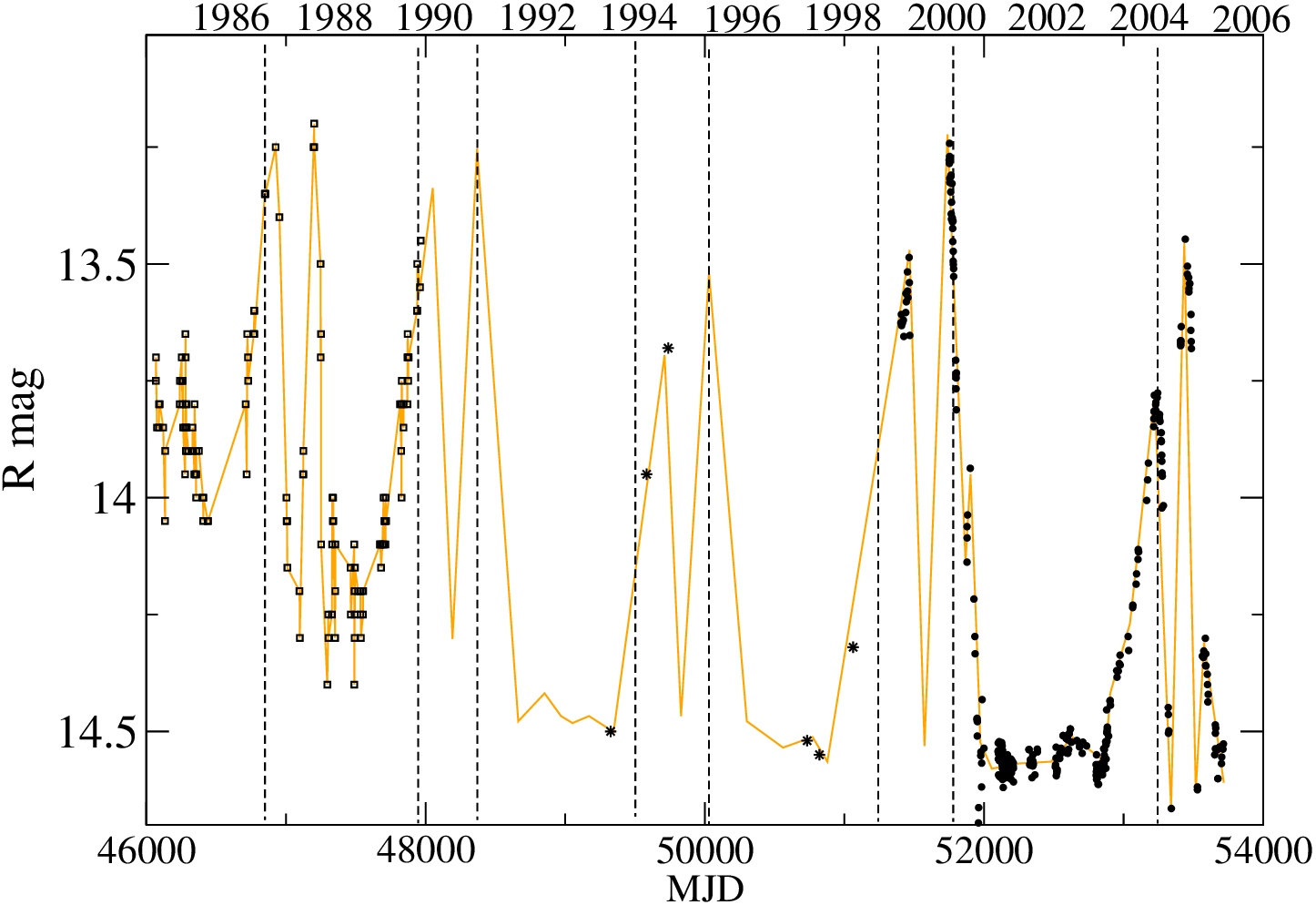}}
\caption[]{Our interpretation of the long-term light curve of \cas. Marked
points represent real observations. Low states occur in between pairs of 
X-ray outbursts, high states after the first outburst.}
\label{inter}
\end{figure}

The low state would correspond to the complete loss of the disc or to a
highly debilitated disc.   During the low state, the source shows the
weakest magnitudes, the bluest colours and the smallest \ha\ equivalent
widths (\ew).  The source exhibits little variability, with changes in the
photometric bands of less than 0.05 magnitudes. Orbital modulation is
detected. Since the source stays in the low state for extended periods (a
few years), the term 'extended low state' (ELS) is used \citep{roch93}.
During the low state the source is X-ray quiet. A disc-less state
occurs when the \ha\ line displays an absorption profile. According to our
interpretation of the long-term light curve shown in Fig.~\ref{inter},
\src\ would have gone through ELS during  MJD 48600--49300 (1991-1993), MJD
50300--50900 (1996-1998) and MJD 52000--52800 (2001-2003).

The issue of whether the disc completely vanishes during the low state or
there still exists some residual emission from a highly debilitated disc is
a very important one, since in the case of total loss one can decouple the
disc emission from that of the central star and hence determine the
astrophysical parameters of the underlying star without any interference
from the disc (Paper I). Disc-less states in \src\ occurred in 1997 and
2001 as the \ha\ line appeared in absorption. Further support for the
complete disappearance of the disc in 2001 comes from the fact that the
standard deviation of the photometric magnitudes during this state does not
follow any trend as a function of wavelength (see Sect.~\ref{pho}). Disc
activity is expected to affect the redder magnitudes more than bluer
magnitudes, hence making the IR magnitudes appear more variable.  Also, 
from the power spectral analysis we observed that when the disc is present
a beat period between the orbital period and presumably the precession
period of the disc was found. Then the fact that no such beat frequency is
detected during the ELS but only a low-amplitude ($\sim 0.01$ mag)
modulation coinciding with the orbital period seems to indicate the absence
of a disc. The 1992 low state would have not been accompanied by the
complete loss of the disc as no absorption profile was attained.

The neutron star spins down during the low state. Previous studies
have reported spin-up episodes throughout the duration of the X-ray
outbursts \citep[see e.g][]{tamu92}. However,  the historical record of the
spin period hardly shows evidence for variability (see Table~\ref{his}). 
Thus, if the neutron star spins up during the outburst, it must spin down
in quiescence.

Disc-less phases have been seen in a number of Be/X-ray binaries but only
two have been observed frequently enough as to allow a meaningful
comparison with \src, namely, A~0535+26 \citep{lyut00,lari01,clar99,haig04}
and X~Per \citep[][and references therein]{clar01}.
The three systems show extended low states, although its duration differs:
$\sim$600 days (MJD 50700--51300) for A~0535+26 \citep{haig04,zait05},
$\sim$800 days (MJD 52000--52800) for \src\ (Fig.~\ref{phot}) and
$\sim$1300 days (MJD 47700--49000) for X~Per \citep{clar01}. 
The long-term photometric light curve of X Per is very similar to that of \src\ 
(cf. Fig~7 in \citet{clar01} with Fig~\ref{phot}), whereas
A~0535+26 exhibits a more choppy behaviour \citep[see Fig.1 in][]{clar99}.

\subsubsection{The high state}

As the disc reforms, it first behaves like the disc of an isolated Be star,
growing in size. This phase is relatively smooth and it is characterised by
a gradual increase in brightness and \ew. After the disc has reached its
maximum size, further mass loss will result in an increase of the disc
density, which can even become optically thick at IR wavelengths. At this
point, the contribution of the disc to the optical/IR photometric
magnitudes and colours is substantial. Fast ($\sim$100 days) and large
amplitude variations (upto 1.5 magnitudes) are seen. The source brightness
alternates between low-state values and optical maxima.  In this state the
shape of the \ha\ line alternates between single-peak and double-peak
profiles, indicative of a perturbed disc.

X-ray outbursts always occur during the high state. They are preceded by
the brightening of the optical and infrared emission. The \ew\ reaches a
relative maximum at the time of the X-ray peak. There seems to exist a
triggering value of the \ew, around 10 \AA, for the onset of the X-ray
outburst. The \ew\ prior to the start of the second outburst is always
larger than that of the first one.  Each X-ray outburst is then followed by
a decrease in the optical/IR brightness but not always by the loss of the
disc. Note that the \ew\ remained fairly unaltered after the 1994, 1998 and
2004 X-ray outbursts (Fig.~\ref{long}). 

When the X-ray outburst does not lead to a low state (i.e. disc loss) the
\ew\ presents a relative minimum a few months after the outburst 
(Table~3, see also Table 1 in Paper~II), as if the disc
would have started a dispersal phase that never materialises. If this
disc-loss phase does not take place, then another X-ray outburst occurs 
about 1--1.5 years later and is promptly followed by the complete
disappearance of the disc. Indeed,  a very interesting feature of the
long-term light curve of \src\ is the fact that, for the last two decades,
X-ray outbursts appear to have occurred in pairs, with two major outbursts
happening in every cycle of disc build-up and loss (Fig.~\ref{long}). 

Although a comparison of the strength of the outbursts is hampered by the
different energy ranges covered by the instruments that detected the
events, we observe that the X-ray luminosity of the first outburst tends to be
larger than the second one (Table~\ref{his}). In one occasion (in 1996, MJD
50300-50400) the second X-ray outburst was followed by a series of orbital
modulated minor outburst (so-called type I) \citep{negu98}. 

The second outburst of the pair, however, is not always detected. This
seems to be the case in the last cycle (2000-2005). Photometric magnitudes
are back to low state values and 1.5 years have passed since the September
2004 outburst. Our latest spectra taken on June 2006 give values of the
\ew\ of $\sim -0.4$ \AA\ and an absorption profile \, with two weak emission
shoulders, a feature typically seen in small disks.
This, however, does not necesarily mean that \src\ has returned to the low
state. In Paper~II we showed that fast transitions between single-peak and
shell profiles are indicative of a precessing warped disc. Another missing
outburst occurred during the  1985-1990 cycle \citep{mend91}. The
similarity between the 2003-2006 and 1986-1989 light curves is striking.
The second optical eruption in 1988 was not accompanied by an X-ray
outburst either. Therefore we suggest that the intervals MJD 46700--48600
(1986-1991) and MJD 52800-present (2003-present) should be considered as
extended high states. Extended high states would occur in association with
missing X-ray outbursts. Normal high states would last for about 3 years
--- the intervals  MJD 47600--48600 (1989-1991), MJD 49300--50300
(1994-1996) and MJD 50900--52000 (1998-2000) would correspond to high
states --- while extended ones would cover the entire cycle of 5 years. In
this case no disc-less phase is achieved.




\subsection{The viscous decretion disc model}

The long-term variability of \src\ can be summarised as follows: the
equatorial disc around the Be star dictates the state of the source. The
quasi-cyclic variations coincide with the time scales for disc build-up and
loss. X-ray outbursts occur in pairs, although the second one may be
missing. In between pairs of outbursts the optical/IR emission and
variability is highly suppressed, while in between the outbursts of a given
pair the optical/IR emission is highly variable.

This behaviour can be explained within the framework of the
truncated decretion disc model \citep[Paper~I, ][]{okaz01,okaz02}.
Initially, the disc grows gradually like in isolated Be stars. However, in
Be/X-ray binaries the surface density of the disc increases more rapidly
that that of isolated Be stars, as a consequence of truncation.  The
truncation of the disc is the result of the resonant torque exerted by the
neutron star, which removes angular momentum from the disc. \citet{okaz02}
locates the truncation radius of \src\ at $0.36-0.39a$, where $a$ is the
semi-major axis of the orbit, for a wide range of values of the viscosity.
The disc becomes optically thick at IR wavelengths and unstable
to radiation-driven warping. Eventually, the disc begins to warp, tilt and
precess.

The photometric variations observed during the high states (1987-1989,
2003-2005), characterised by large ($\simmore$ 1 mag) and relatively fast
($\sim$100 days) amplitude variations, are rather large when compared to
typical values seen in isolated Be stars \citep{dach88}. In addition, the
brightest magnitudes are accompanied by historic maxima of the \ew. These
extreme values of the \ew\ correspond to line shapes much narrower than at
other times. In the framework of the disc deccretion model, the coincidence
of very high values of the \ew\ with narrow, single-peaked lines and very
bright optical and infrared magnitudes during the high state, leads to the
conclusion that extreme values happen because the disk becomes warped and
occasionally presents a much higher surface to the observer (see Paper~II).
Therefore the changes in magnitudes should be the consequence of two
different effects: on the one hand, an increase in the disk size or
density; on the other hand, a change in the emitting surface observed. They
should not be interpreted as fast phases of disc loss and reformation,
despite the fact that the photometric magnitudes fall to ELS values.
This strongly disturbed phase would begin after the first outburst.

One possible explanation for the difference between the first and second
outburst could be the different state of the disc, with  the first one
occurring when the disc is still in a quasi-stable state while the second
one when the disc has been distorted. The distortion of the disc leads to
the interaction with the orbiting neutron star. If the interpretation of
the 22.15 day period (Sect.~\ref{pow}) as the beat period between the
orbital period and precession is correct, then the time scales associated
with the perturbed disc ($\sim$250 d) would be a factor $\sim$2 longer than
the duration of the optical eruption associated with the missing X-ray
outburst ($\sim$150 d). In this context the missing outburst would simply
reflect the fact that no interaction between the neutron star and the
distorted disc took place during the $\sim$6 orbits that the optical
eruption lasted.

An alternative explanation for the missing X-ray outburst requires the
formation of an accretion disc around the neutron star \citep{mend91} and
it is based on the centrifugal inhibition at the edge of the neutron star
magnetosphere \citep{stel86}. In the 1988 and 2005 events the matter
accumulated in the disc would not be enough to overcome the propeller
effect. 

The $(B-V)$ colour index appears as a good indicator to distinguish between
the two outbursts. While the $(J-K)$ colour (also $(V-R)$ and $(R-I)$, not shown
in Fig~\ref{phot}) follows smoothly the state of the disc, $(B-V)$ is rather
insensitive to photometric or spectral changes. Only prior to the
destructive outburst of September 2000 does it display high amplitude
variability (Fig.~\ref{phot}).

In addition to the missing outbursts, the fact that the  first X-ray
outburst does not modify, at least dramatically, the strength of the \ha\
line also needs to be explained. In this respect, one may wonder how much
of a disruption an X-ray outburst represents. As an order of
magnitude estimate, we can derive the mass content in the disc simply as

\[M=\int{\rho\, dV}\]

\noindent with $\rho=\rho_0(R_*/r)^{n}$ and $dV=2\pi rH dr$. Here $\rho_0$
is the density and the inner radius of the disc (at $r\sim R_*$), $R_*$ the
star radius,  $H$ the disc height and $n$ an exponent defining the
density law. According to \citet{wate88} $n$ varies in the range 2-4 in
most BeX. The typical radius of a B0.2 star, is $R_*=8\rsun$ and assuming 
typical values for the disc radius $R_{\rm d}=5R_*$, and inner density
$\rho_0=10^{-10}$ g \citep{telt98} cm$^{-3}$ and $H=0.03R_*$ \citep{neok01}
the mass is estimated to be $2.6 \times 10^{-9}$ $\msun$ for $n=2$ and $7.8
\times 10^{-10}$ $\msun$ for $n=4$.

On the other hand, a mean X-ray luminosity of $10^{37}$ erg s$^{-1}$ for
about a month requires the transfer of $\dot{M}=L_x(R/GM)=5.4 \times
10^{16}$ g s$^{-1}$ or $7 \times 10^{-11}$ $\msun$ month$^{-1}$, that is
just $\simless$ 10\% of the mass in the disc. This relatively small effect
on the mass content of the disc would explain the fact that the strength of
the \ha\ line does not change much after the first outburst 
\citep[see also][]{nort94}.

\section{Conclusion}

We have presented the results of our monitoring of the Be/X-ray binary
\src/\cas. Our 2000-2006 data represent the most complete optical and
infrared photometric study of \src/\cas\  made up to now. The combination
of our observations over the last decade with published data allowed us to
investigate the correlation between the optical, infrared and X-ray
emission.  X-ray outbursts, which come into pairs, occur when the source is
photometrically bright and the \ha\ line appears strongly in emission. The
equatorial disc around the Be star dictates the state of the source: if the
disc is absent the source is in the low state. In the high state large
amplitude variation and asymmetric spectral lines denotes a disturbed disc.
The time scale for loss and build-up of the equatorial disc is 5 years,
approximately 3 of which correspond to the high state and 2 to the low
state. However, this quasi-periodic behaviour is broken if one of the X-ray
outburst is missing. In this situation the high state covers the entire
cycle. Bright magnitudes are observed because the disc warps and presents a
larger surface to the observer.  The fact that the infrared and optical
magnitudes and the \ha\ equivalent width do not change dramatically after
the first outburst may indicate that, at the time of the outburst, the disc
had not been distorted yet. It is after the first outburst that the disc
becomes unstable, warps and tilts. The second X-ray outburst takes place
during this phase of strongly disturbed disc. The disc finally disappears,
partly reabsorbed by the B star and partly used up to power the X-ray
outburst.

\begin{acknowledgements}

We thank A. Manousakis, A. Di Paola and M. Dolci for their help in
some of the observations. IN is a researcher of the programme {\em Ram\'on
y Cajal}, funded by the Spanish Ministerio de Educaci\'on y Ciencia (MEC)
and the University of Alicante, with partial  support from the Generalitat
Valenciana and the European Regional Development Fund (ERDF/FEDER). This
research is partially supported by the MEC under grant AYA2005-00095. VL
and AA acknowledge hospitality of the Rome and Teramo observatories during
the observations in Campo Imperatore. This work has greatly benefited from
the ING service programme. Skinakas Observatory is a collaborative project
of the University of Crete, the Foundation for Research and
Technology-Hellas and the Max-Planck-Institut f\"ur Extraterrestrische
Physik. The near-infrared data were obtained with AZT-24 telescope
operated  under agreement between Pulkovo, Rome and Teramo observatories.
The INT and JKT are operated on the island of La Palma by the Isaac Newton
Group in the Spanish Observatorio del Roque de Los Muchachos of the
Instituto de Astrof\'{\i}sica de Canarias. This research is partially based
on data from the ING Archive. Part of the data presented here have been
taken using ALFOSC, which is  owned by the Instituto de Astrof\'{\i}sica de
Andaluc\'{\i}a (IAA) and operated at the Nordic Optical Telescope under
agreement between IAA and the NBIfAFG of the Astronomical Observatory of
Copenhagen. Partially based on observations obtained at the Observatoire de
Haute Provence (CNRS, France), observations obtained at the Asiago
observatory and the Centro Astron\'omico Hispano Alem\'an (CAHA) at Calar
Alto, operated jointly by the Max-Planck Institut für Astronomie and the
Instituto de Astrof\'{\i}sica de Andaluc\'{\i}a (CSIC). This research has
made use of NASA's Astrophysics Data System Bibliographic Services and of
the SIMBAD database, operated at the CDS, Strasbourg, France. 

\end{acknowledgements}

\end{document}